\documentclass{iaus}
\usepackage{graphicx}

\title[Activity diagnostics and indices]
{Solar and Stellar Activity:\\ Diagnostics and Indices}

\author[P. G. Judge and M. J. Thompson]
{Philip G. Judge$^1$ and Michael J. Thompson$^1$}

\affiliation{$^1$High Altitude Observatory, National Center for Atmospheric Research \\ PO Box 3000,
Boulder CO 80307-3000, USA\\ email: {\tt judge@ucar.edu,\  mjt@ucar.edu} }

\pubyear{2012}
\volume{286}  
\pagerange{???-???}
\jname{Comparative Magnetic Minima: Characterizing quiet times in the Sun and stars}
\editors{C. Mandrini and D. Webb, eds.}
\begin{document}

\maketitle

\begin{abstract}
We summarize the fifty-year concerted effort to place the ``activity''
of the Sun in the context of the stars. As a working definition of
solar activity in the context of stars, we adopt 
those {\em globally--observable} variations on time scales below 
thermal time scales,  of $\sim 10^5$ yr for the convection zone.  So defined, activity 
is dominated by magnetic--field evolution, 
including
the 22--year Hale cycle,
the typical time it takes for the quasi-periodic reversal in which the
global magnetic--field takes place. This is accompanied by sunspot
variations with 11 year periods, known since the time of Schwabe, as
well as faster variations due to rotation of active regions and flaring.
``Diagnostics and indices'' are terms given to the {\em indirect}
signatures of varying magnetic--fields, including the photometric (broad-band)
variations associated with the sunspot cycle, and variations of the
accompanying heated plasma in higher layers of stellar atmospheres
seen at special optical wavelengths, and UV and X-ray wavelengths. Our
attention is also focussed on the theme of the Symposium by examining
evidence for deep and extended minima of stars, and placing the 70--year
long solar Maunder Minimum 
into a stellar context.
\end{abstract}

\firstsection 
\section{Motivations: Do we remotely understand the Sun?}

Globally, the Sun appears to a remarkably constant
star.  RMS variations in irradiance (observed bolometric flux), as measured from Earth
orbit since 1978, are
$\approx 0.04$\%. The theory of stellar structure and atmospheres has
reached a remarkable level of agreement with critical measurements, 
now including the remarkable tool of
astero-seismology of solar-like stars. While there remain important debates, such as
the apparent disagreement between solar interior abundances derived from helioseismology and surface abundances derived
from spectroscopy, 
we have nevertheless gained confidence that 
our basic theory and 
understanding has withstood many onslaughts and experimental
challenges. A particular success is the resolution of the ``solar neutrino
problem'' in terms of particle, not solar physics. 

The above statements might by some be considered a reasonable
summary of solar physics.  The global and long time scale
solar behavior might even be considered a ``dead'' subject for most
astrophysicists, were it not for one inconvenient fact.  The 0.04\% 
RMS variations occur on time scales of decades and less,  orders of magnitude
smaller than the thermal relaxation (Kelvin-Helmholz) timescale of $10^5$ years for the
Sun's convection zone. This is a further 4 orders of magnitude smaller than the
``diffusion time'' for global magnetic--fields. Remarkably, overwhelming
evidence indicates that the solar magnetic--field, evolving {\em globally on
decadal time scales}, is the culprit. 

The global Sun is well described by magneto-hydrodynamics (MHD), in which equations
of hydrodynamics are coupled to 
an equation for magnetic--field
evolution (the ``induction equation'') because the solar plasma is
highly conducting.  
The system of MHD equations 
is highly non-linear; fluid motions generate electric currents which generate magnetic--fields; 
magnetic--fields with the electric currents act through the Lorentz
force on the fluid, and so forth.  Because of enormous physical
scales, inductance effects dominate the electromagnetic--fields.
Because the plasma is highly conducting, steady electric fields are
essentially zero; all the EM energy lies in the magnetic--field.  
By radius, the outer 30\% of the Sun is fully turbulent, as radiation is unable to carry the 
energy flux and thermal convection takes over.  Just beneath the observed photosphere, 
turbulent fluid motions carry all of this energy flux.   Under these conditions 
we would expect that 
the Sun would exhibit 
the rich landscape of non-linear phenomena, including chaotic behavior.  Such 
phenomena are of course observed
in the form of small scale, dynamic granulation.  However, 
when we view the behavior of the global 
solar magnetic--field in this fashion, several questions come to mind. For
example, why should the Sun's magnetic--field appear so prominently in
the form of intense concentrations- sunspots?  Why does the solar cycle have so much
{\em order} (quasi-cycling behavior; Hale's polarity law, Joy's law of
tilt of sunspot bipoles, active longitudes)?    Why does the global field reverse 
every 22--years?  These are profound, unanswered questions 
of solar physics, related in some way to the order imposed by the
(differential) solar rotation (Parker 1955).   When viewed in terms of first principles, 
such ordered behavior  is surely {\em unexpected}.  As we will see below,
many solar-like stars do not exhibit this level of order. 
In this sense, some components of solar magnetic variability, such
as  extended and Maunder-like minima studied at this Symposium, are 
just some examples of stochastic behavior in our limited 
historical record of our non-linearly varying star.  

To add insult to injury, 
the extent of our ignorance the Sun's variable magnetism 
is highlighted by 
recent sobering results.  Brown et al. (2010)  made numerical experiments of ``rapidly rotating Suns''. 
General consensus was that long-lived, ordered fields in stellar
interiors, needed to explain the order in sunspot behavior,  should exist should exist mostly outside
convection zones.  Yet Brown and colleagues found 
coherent ``wreaths'' of magnetic--field living entirely {\em within} highly turbulent convection
zones, 
for many convective turnover times.  
Further, in the abstract of 
Brown et al. (2011), we read
\begin{quotation}
  ``Striking magnetic wreaths span the convection zone and coexist with
  the turbulent convection. A surprising feature of this
  wreath-building dynamo is its rich time dependence. The dynamo
  exhibits cyclic activity and undergoes quasi-periodic polarity
  reversals where both the global-scale poloidal and toroidal fields
  change in sense on a roughly 1500 day timescale. These magnetic
  activity patterns emerge spontaneously from the turbulent flow..''
\end{quotation}

Given this state of affairs, 
we review solar-stellar research to shed light on these basic
issues, in special relation to states of minimum
magnetic activity, such as the Sun's recent extended minimum and the Maunder
Minimum.  The stars offer the opportunity to ``run the solar experiment again'', 
with the caveats that (1) no two stars are identical, and (2) that we only observe the Sun from 
our special viewpoint in the ecliptic plane, only 7$^\circ$ from the
solar equatorial plane.

\section{The need for diagnostics and indices}

``Diagnostics and indices'' -- proxies for magnetic activity on
solar-like stars -- are required because ``direct'' measurement of
magnetism of the Sun-as-a-star is difficult.
Measurements of stellar magnetic--fields are based almost
exclusively upon the polarization of spectral lines induced through the 
Zeeman effect, or the increased width of certain Zeeman-sensitive lines,
because most lines are not fully split (Zeeman splitting is less than 
linewidths).  
Hemisphere-integrated 
Zeeman signals for stars of {\em
solar type} (mass, age, rotation rate) 
are particularly difficult to measure. 
Considering the Sun itself, 
the peak-to-peak, disk-integrated variation of polarized light 
from the kilo-Gauss sunspot fields varying with
the solar cycle is estimated to be 
equivalent to a mere 2 Mx~cm$^{-2}$ average flux density
(Plachinda \& Tarasova 2000).
For typical spectral lines at visible wavelengths, this leads to 
a tiny polarization (a few times $\sim 10^{-4}$)\footnote{
If the Sun were observed nearly pole-on,
the solar cycle might be seen with a higher amplitude as the poles 
are dominated by large areas of magnetic--field of the {\em same}
polarity whose flux density varies with an average amplitude of $\sim
10-20$ Mx~cm$^{-2}$ Schrijver \& Harvey(1994).}.

It is important to remind oneself that 
remotely sensed ``magnetic--field strengths'' (in units of
Gauss or Tesla)
through spectral lines, even in the Sun, are direct measurements
only when spectral lines are fully split (Zeeman splitting $>$ 
line widths, usually dominated by Doppler broadening).  
So direct measurements are possible for very intense fields and 
at longer (infrared) wavelengths, but generally this is not the case
in practice. 
Solar ``magnetograms'' generated by ground- and spaced- based instruments
exclusively work in the unsplit regime, where the first order 
polarization signature 
is proportional to $\int \bf{B}\cdot d{\bf S}$, ${\bf S}$ being a
vector along the line of sight, observed with pixels of projected area $S$. Thus if 
one has a field- (and polarization-) free surface of area $S$ and a single magnetic
structure with line-of-sight strength $B_\parallel$ occupying 
an area $s < S$, one measures an average ``flux density'' $
\frac{s}{S} B_\parallel$~ Mx~cm$^{-2}$,  where the actual field strength is $B$ G.
This difference between the kG field strengths of sunspots and
the average surface flux density in Mx~cm$^{-2}$ in unresolved
solar-like stars makes the disk-integrated net polarization small.

The magnetic polarization of Sun-as-a-star spectra is also limited for other 
physical reasons.  Sunspot groups are small compared with stellar
hemispheres: the absence of magnetic monopoles means that opposite
polarities appear together in spot pairs, in this case the dominant
(first-order) Zeeman induced polarization almost cancels in the
integrated light (the same flux emerges in one polarity as returns
through
the other polarity in a given active region). 
Force balance in the photosphere limits the field
strengths to near-equipartition values where $B^2/8\pi < 
\frac{3}{2}nkT$, a few thousand G.  Thus, 
stellar Stokes $V/I$ measurements are intrinsically very weak. 
Few solar-like stars have been targeted using polarimetry, and 
those which have are not really of solar type: they are younger, 
more rapid rotators.

This boils down to the {\em necessity}, for all but a few special
(rapidly rotating and active) 
targets, to look for other signatures of magnetic
activity. Hence those ``diagnostics and
indices'' of our  title: the variable radiation in well-known
chromospheric (optical, UV lines), transition region (UV) and coronal
(EUV/X-ray) features. These are important because they 
correlate with spatially resolved magnetic structures measured on the
Sun 
(e.g., Schrijver \& Zwaan 2000),
they vary considerably with the solar sunspot cycle, and, being
radiators of dissipated
magnetic energy ($B^2/8\pi$), they are not subject to
the cancellation of signals arising from 
opposite polarity fields on the visible
hemisphere.   Recently, helio- and astro- 
{\em seismology}  have become important additions 
to the toolkit for activity indicators.  

\begin{figure}[ht]
\begin{center}
\includegraphics[width=5.6in]{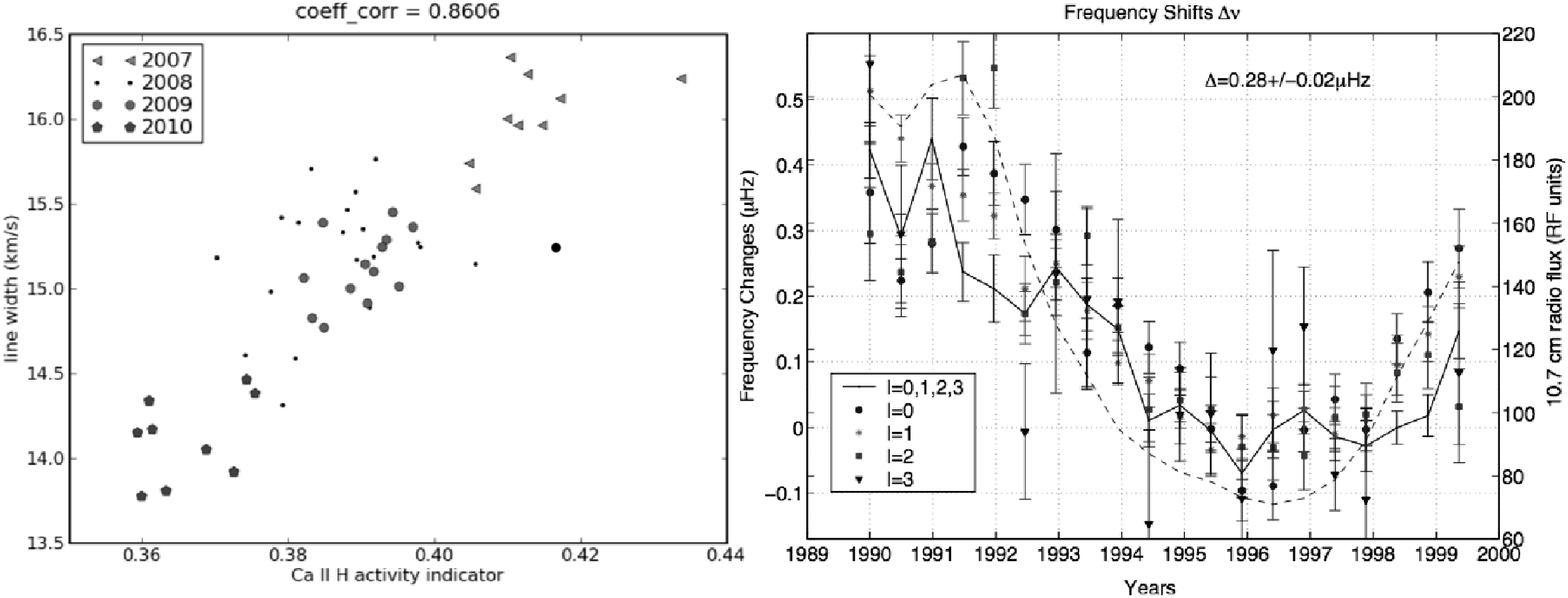} 
\caption{Left panel: variations in a Zeeman-sensitive 
line width (ordinate) versus chromospheric 
emission index (abscissa), for the young and active solar-like G~8 V  star
$\xi$ Boo A (Morgenthaler et al. 2010).   Right panel: 
solar measurements acquired as if the Sun were a star, showing
relative mode oscillation frequency differences versus time. The 10.7 cm radio flux,
a well-known magnetic activity index  forming through free-free
emission near the coronal base, is
plotted as a dashed line 
(Salabert et al. 2003). 
}
  \label{fig:morgen}
\end{center}
\end{figure}
The correlation between Zeeman signals of magnetic--field
and the classical ``Ca II'' (chromospheric brightness)
index has been demonstrated clearly for at least one solar-like star,
other than the Sun itself (Fig.~\protect\ref{fig:morgen}).  The star, $\xi$ Boo A of spectral type 
G8 V, with 
rotation period $\sim 6$ days, is significantly younger and more
active than the Sun.   The figure
also shows a correlation between low-$\ell$ oscillation frequencies
and the F10.7 radio flux, both measured for the Sun-as-a-star, from
Salabert et al (2003). 

In summary, {\em for Sun-like stars}  we must rely on the fact that
stellar magnetic--fields are correlated with
variations in stellar optical photometric measurements (both
in global helioseismic techniques and in photometry revealing the
passage of spots and plages across stellar disks), in UV fluxes
and in X-ray fluxes, through their presumed connection to the magnetic features - 
spots, plages and network - which are directly seen  on the Sun. 
There is a large literature on these relationships for the Sun, both seen-as-a star
and for features across the solar disk. 
Much material is nicely reviewed by {Schrijver \& Zwaan (2000).

\section{Comparing magnetic minima in the Sun and stars}

\subsection{Practical considerations}

Systematic observations of stellar magnetic activity began in 1966,
when 
Wilson (1968) began the ground-based ``Mt. Wilson survey''.   
Table~\protect\ref{tab1} 
 lists significant programs that have contributed to the database
of measurements of magnetic activity on decade and longer time
scales.  The table includes the important Fraunhofer Ca~II ``H and
K'' lines at 397 and 393 nm, whose line cores form in 
chromospheric plasmas.  
\begin{table}[h]
  \begin{center}
  \caption{Significant synoptic observational programs of solar
   and stellar magnetic variability} 
  \label{tab1}
 {\scriptsize
  \begin{tabular}{lllll}\hline
Target & Program & dates & Observable & notes\\
\hline 
The Sun \\ 
&Various$^a$ & 1608-&  Sunspot counts & $\sim$ daily \\ 
&Ottawa$^a$      & 1947- & F10.7 (10.7 cm radio flux) & daily\\ 
&Sacramento Peak$^b$  & 1974- &Solar disk Ca~II &$\sim$ daily \\ 
&Kitt Peak$^b$     &  1974 & Solar disk Ca~II &$\sim$ daily \\
&Various spacecraft$^c$     & 1978- & total irradiance & daily \\  
&Various spacecraft$^d$  & 1981- & IR-X-ray spectral irradiances & daily
\\
{Stars} \\
&Mt. Wilson$^e$  & 1966-1996 & Ca~II field G\&K stars   & daily/seasonal  \\
&    Lowell/Fairborn$^f$ & 1984- & Stromgren $b,y$ colors & daily/seasonal \\  
&HAO SMARTS$^g$  & 2007- &  S. hemisphere field G\&Kstars & weekly/seasonal \\ 
{Sun and stars} \\ 
& SSS/Lowell$^h$  & 1994-  & Ca~II field G\&K stars, integrated sunlight & weekly/seasonal\\
\hline 
 \end{tabular}
  }
 \end{center}
\vspace{1mm}
 \scriptsize{
 {\it Notes: 
(a) See, for example, Hufbauer (1991).  
(b) 
Livingston et al. (2010).
(c) Frohlich (2011). (d) Rottman (2006).
(e)  Baliunas et al. (1995).
(f)   Radick et al.  (1998). 
(g) Metcalfe et al.  (2010). 
(h) Hall (2008). 
}\\
}
\end{table}

The Ca~II lines are important since they represent the longest
continuously observed diagnostic of stellar surface magnetism, and 
they contain data both on rotation and dynamo action on stars.   
A traditional ``index'' is the ``S-index'', which measures 
the line core (= chromospheric component) integrated over
a triangular filter, relative to the neighboring ``continuum'',
thereby
giving a normalized measure of the chromospheric to 
broad-band flux.  The S index will suffice for our use below, 
but the reader should be aware that others are in use, some of
which are more closely related to basic stellar parameters.  This
issue is
discussed, for example, by Hall (2008).

A fundamental difficulty in 
comparing stellar and solar activity indices  is 
that it is a ``bandwidth--limited'' exercise. At best, 
any star has been observed for 45 years on a daily basis, subject
to seasonal observability constraints.   If in an ensemble
of such stars we truly had {\em identical} suns with their rotation axes,
like the Sun, just a few degrees from the plane-of-the-sky, then we
could invoke ergodicity and compare the Sun's statistical 
behavior in time
with the variations seen among the stellar sample.  Unfortunately,
this exercise is made complicated by significant 
dependences of the magnetic indices on stellar mass, age, metallicity,
and orientation of the rotation axes.   There is no solar `` identical
twin'', so we
cannot yet exactly ``rerun the solar experiment'' under controlled
conditions, and some 
care is needed\footnote{This is an area we expect future progress 
from astroseismology, see section \protect\ref{sec:seismology}.}.  

Now if  we had observed the Sun daily 
for 45 years, we would have captured 2 complete Hale magnetic 22--year
cycles, unless we had been observing during the Maunder Minimum 
(see Fig.~\protect\ref{fig:ssn}).  
\begin{figure}[ht]
\begin{center}
 \includegraphics[width=4.4in]{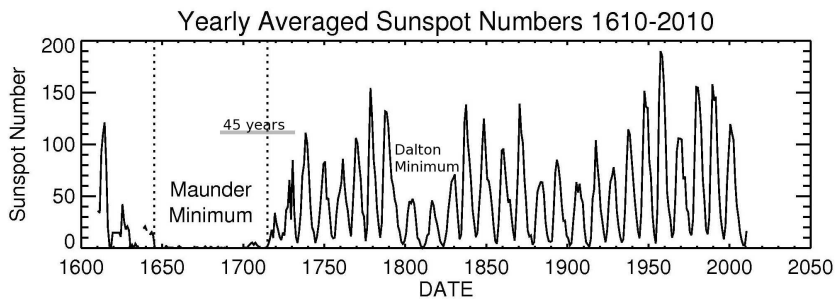} 
 \caption{Yearly averaged sunspot numbers.  The bar marked 
``45 years'', placed arbitrarily near the end of the Maunder Minimum,  shows the span of time for which we have continuous 
observations of magnetic activity in stars.  
}
  \label{fig:ssn}
\end{center}
\end{figure}
Importantly, if we place a 45 year window across 
any particular period in the sunspot record, it is clear that at most we
would
observe four minima in the sunspot cycle.  Further, if we were to 
observe the Sun somewhere near the Maunder Minimum, we might
conclude that the Sun is in not a regular cycling state, but instead
is perhaps a ``flat activity'' or even ``irregularly varying'' star.  
The one thing that we would expect, though, is that the {\em mean
  value}
of the chromospheric activity index (the Ca II ``S'' index for
example) is near the low end of
the stellar distribution.  This is because
solar ``S-index'' data were obtained
during the last 40 years or so of high sunspot numbers
(Fig.~\protect\ref{fig:ssn}) and yet they are relatively low in
a stellar context (Fig.~\protect\ref{fig:bal1}). 

Henceforth we will therefore discuss 
{\em relatively inactive stars, which are also slow
rotators} 
e.g. Noyes et al.  (1984a).   Techniques 
requiring strong
magnetic--fields in which different spot polarities are spectrally 
separated by differential Doppler shifts in 
rapidly rotating stars, such as Zeeman Doppler
Imaging Semel (1989), cannot therefore be applied.
Lastly, even to examine ``comparative minima'', the stars must be
\begin{enumerate}
\item “Single” (weak star-star/ star-exoplanet interactions)
\item Cycling, or
\item In a GM state (“flat”)
\end{enumerate}
Mostly we will be restricted to single stars similar in mass,  age and (hence) activity to the Sun. 

\begin{figure}[ht]
\begin{center}
 \includegraphics[width=5.4in]{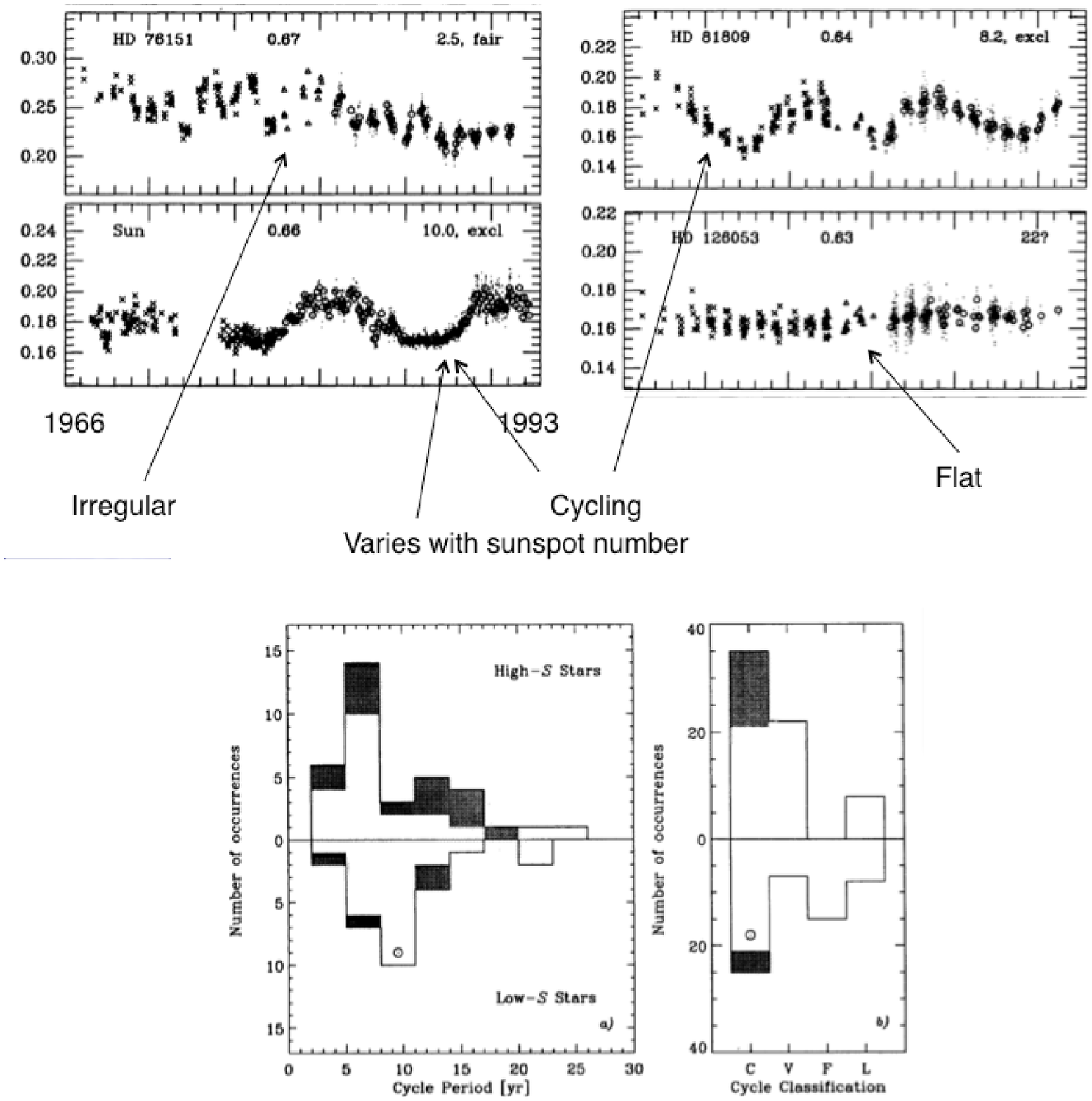} 
 \caption{A selection of different stellar behavior measured in the
   Mt. Wilson survey up to 1995, reported by
   Baliunas et al.  (1995).   The upper four panels show representative 
time series of the ``S-index'', the lower panels the number of stars
as a function of cycle period, and the number of stars which were
classified as cyclic (C, like the current Sun), irregularly variable
(V), flat (F, the Maunder Minimum Sun?) and
with a long-term trend (L).  
}
  \label{fig:bal1}
\end{center}
\end{figure}

\subsection{The stellar records -- ``settled'' issues}

Analysis of histograms such as those shown in  
Fig.~\protect\ref{fig:bal1}  suggests that 
the Sun is a typical 
``low-S'' star.  In the stellar Ca II data, almost
as many such stars don't “cycle” as do Baliunas et al.  (1995).
Such results are broadly confirmed by the Lowell and other
(non-synoptic, but otherwise relevant) observing programs and campaigns 
(such as a campaign on solar-age stars in M67 by Giampapa et al.  2006).
So, in spite of practical difficulties, the present Sun does appear
broadly to behave like a significant fraction of G and K main sequence
stars, at least according to the Ca~II and broad-band photometric
records.    

Several contentious issues in 
the comparison of the Sun and stars are now largely, if not
completely, 
resolved as we have gained more data and a better understanding 
of existing data.   Most points of debate arose because of 
``small number statistics'', but there are interesting 
biases of a physical origin that have also led to confusion. 

A debate concerning the statistical occurrence of stars in a state
perhaps equivalent to the Maunder Minimum arose soon after a
publication of Baliunas+Jastrow (1990), which showed a
bi-model distribution of chromospheric S-indices which seemed to
correlate
with whether low activity stars were in a cycling or flat state.  It
was suggested that the two distributions corresponded to the Sun in
its cycling versus non-cycling states.   
But the published 
correlation has not survived scrutiny from two perspectives: first, 
Hall \& Lockwood (2004) showed that when S-index data are analyzed 
according to seasonal averages (i.e. equal weights given to 
equal spans of time) the bimodal distribution disappears. Secondly,
Wright (2004) used accurate parallaxes from the {\em HIPPARCOS} 
mission to show that many flat activity stars appear to be
significantly evolved 
above the main sequence.   

Another question arose concerning stellar X-rays and activity cycles.  It
has been known for decades that the variances in solar activity indices
increase with the temperature of the plasma from which they 
originate.  Thus,  the Ca~II index varies more than radiation from 
the cool, dense 
photosphere; vacuum UV radiation from the mid-upper chromosphere
varies more than Ca~II, EUV radiation from well-known 
coronal lines (with no
change in principle quantum number) observed routinely by missions such as
{\em SOHO}, {\em TRACE} and {\em SDO} varies more still, and then 
X- and $\gamma$- rays vary the most.  If we can see stellar cycles in Ca~II 
so clearly (Fig.~\protect\ref{fig:bal1}), we should see them with enormous
amplitudes in X- and $\gamma$- rays.   Yet, for a decade or so, we did
not.

An important feature of X-rays from solar type stars was highlighted
by Schmitt (1997).  Based on a volume limited sample of stars
observed by {\em ROSAT}, he showed that {\em the soft X-ray flux in
  the 0.1-2.4 keV bandpass has a lower limit of $10^4$
  erg~cm$^{-2}$~s$^{-1}$}.  To have such a lower limit seemed to contrast with
solar data from the Soft X-ray Telescope on {\em Yohkoh}, which exhibited
enormous variations in count rates over the solar cycle\footnote{See,
 for example, the ``solar cycle in X-rays'' images which were widely
 distributed among the community, at {\tt
    http://solar.physics.montana.edu/sxt/}}.  Asking the question,
``where are the stellar cycles in X-rays?''
Stern, Alexander \& Acton (2003) computed the soft X-ray irradiance
variations from the {\em Yohkoh} data showing a maximum to minimum
ratio, effectively smoothed with a 4th order polynomial fit, of 30.
Inspection of their figure 3, in which the lowest count rates are
systematically overestimated, suggests this to be a lower limit, more
likely ratios appear closer to 100 and can approach 1000.
They compared these data with soft X-ray data for solar-like stars in
the Hyades group from {\em ROSAT} IPC data.  Discounting unlikely
fortuitous phases linking stellar cycles with the epochs of stellar
observations, they concluded that ``Hyades F-G dwarfs have either very
long X-ray cycles, weak cycles or no cycle at all''.  This puzzle has
since been resolved noting that inter-instrumental calibrations must
be done taking particular care to define the precise response of the different
detectors used for solar and stellar work
Judge, Solomon \& Ayres (2003).  In a result anticipated by
Ayres (1997), Judge and colleagues simulated ROSAT soft X-ray count rates of the
Sun using solar soft X-ray data from the SNOE experiment. They found that through the
0.1-2.4 keV, ROSAT channel, the Sun would would have observed factors
of between 5 and 10, maximum / minimum flux.  Cyclic soft X-ray variations of
a factor of several (max/min) have in fact since been seen in this
channel in the K5 V 
star 61 Cyg A (Robrade,Schmitt \& Hempelmann 2007). The X-ray
variations were in phase with Ca~II emission over a period of 12 years.  

A related problem was the reported ``disappearance'' of the corona
of $\alpha$ Cen~A, a G2 V star similar to but some 20\% larger than the Sun itself
(Robrade, Schmitt \& Favata 2005).  They reported a factor 25 reduction in the
``X ray luminosity'' of this star over two years of observations with 
the XMM-Newton satellite, ostensibly between energies of 0.2-2.0 keV.
This seemed to suggest that the Sun's corona had the possibility
of almost disappearing in a span of 2 years, something unprecedented 
since X-ray data were first acquired some 6 decades ago.   The 
dilemma was resolved when Ayres et al.  (2008)
obtained LETGS spectra with the {\em Chandera} satellite.   The
results, highlighted in Fig.~\protect\ref{fig:ayres}, show clearly
that while the higher energy soft X-rays decrease enormously, the EUV 
transitions now so familiar to us in solar images from SOHO, TRACE and 
SDO, remain strong.   A slight drop in the average coronal
temperature serves to remove soft X-rays from the spectrum,
originating from plasma near 2MK, while at the same time keeping the 
EUV coronal transitions (17.1nm, 19.5nm) strong.   
\begin{figure}[ht]
\begin{center}
 \includegraphics[width=5.4in]{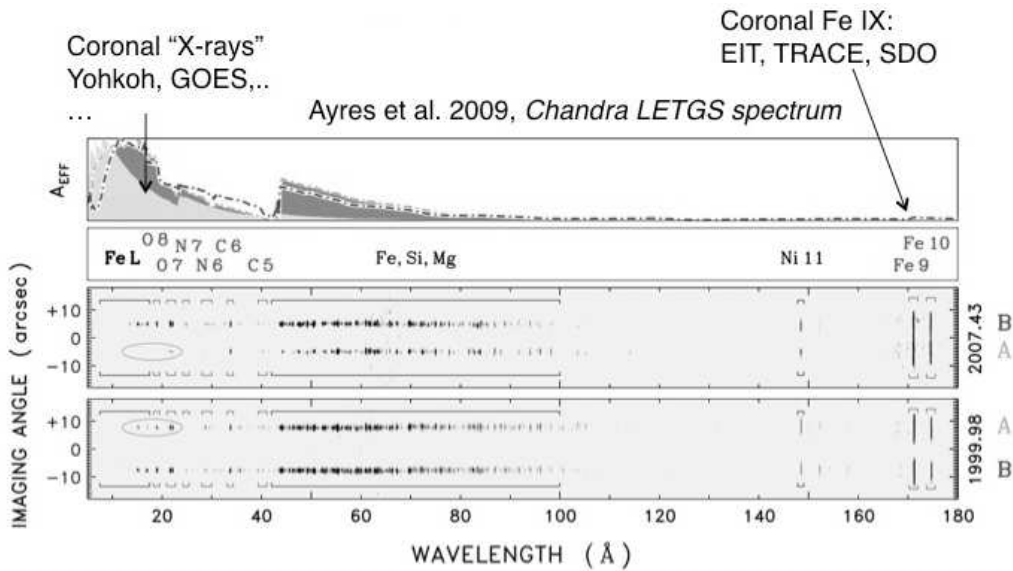} 
 \caption{EUV and soft X-ray spectra of $\alpha$ Cen A and B 
reported by Ayres et al.  (2008). Ellipses highlight the spectral
comings and goings of soft X-rays in $\alpha$ Cen A at two different
epochs, while the EUV data at energies some 10$\times$ lower remain 
similar.}
  \label{fig:ayres}
\end{center}
\end{figure}

In summary, the higher energy behavior of the Sun in the context of
sun-like stars is no longer generally believed by most to be anomalous.
Returning to lower energies, 
it was earlier believed that perhaps the Sun's broad-band ($\propto $
irradiance) variations were lower than other stars for a given
variation in S-index (Radick et al.  1998, Lockwood et al.  2007).
But the evidence that the Sun is anomalous in this sense is not so
clear (Hall et al.  2009).  More sensitive stellar photometry has
revealed stars with lower photometric variability (the Sun's
variability measured since 1978 is close to the detection limit 
for ground-based telescopes), and the nearest
``solar twin'' (18 Sco) has a photometric behavior much like that of the
Sun (Hall 2008).

\subsection{The stellar records- unsettled issues}

It may not be possible to invoke ``small number statistics'' to 
explain the apparently anomalous behavior of the Sun shown 
in Fig.~\protect\ref{fig:bohm}. Using carefully vetted data from
Saar \& Brandenburg (1999), in which stellar S indices were used to
derive both rotation periods and, for those cycling stars, cycle
periods, B\"ohm-Vitense (2007)  plotted the derived cycle periods
against rotation period.  Her motivation was to examine the role of
deep--seated and near--surface shear layers as potential sources for the
re-generation of magnetic--fields (dynamos) on stars, discussed, for
example, by
Durney, Mihalas \& Robinson (1981).   B\"ohm-Vitense proposed that the 
appearance of the two branches (I= ``inactive'' and A='``active'') in the
figure (with some stars plotted twice if they showed two
cycle periods) may correspond to the actions of shear-generated
magnetic--fields beneath the
convection zone (I) and above it (A). The Sun lies
squarely between the two branches.   This kind of result, although
based on a limited stellar sample, is exciting because it
may represent a clear departure of the Sun from two relatively simple
proposed sources for dynamo action in stars.  It is precisely this
kind of disagreement which can lead to advances in our understanding. 

\begin{figure}[ht]
\begin{center}
 \includegraphics[width=2.4in]{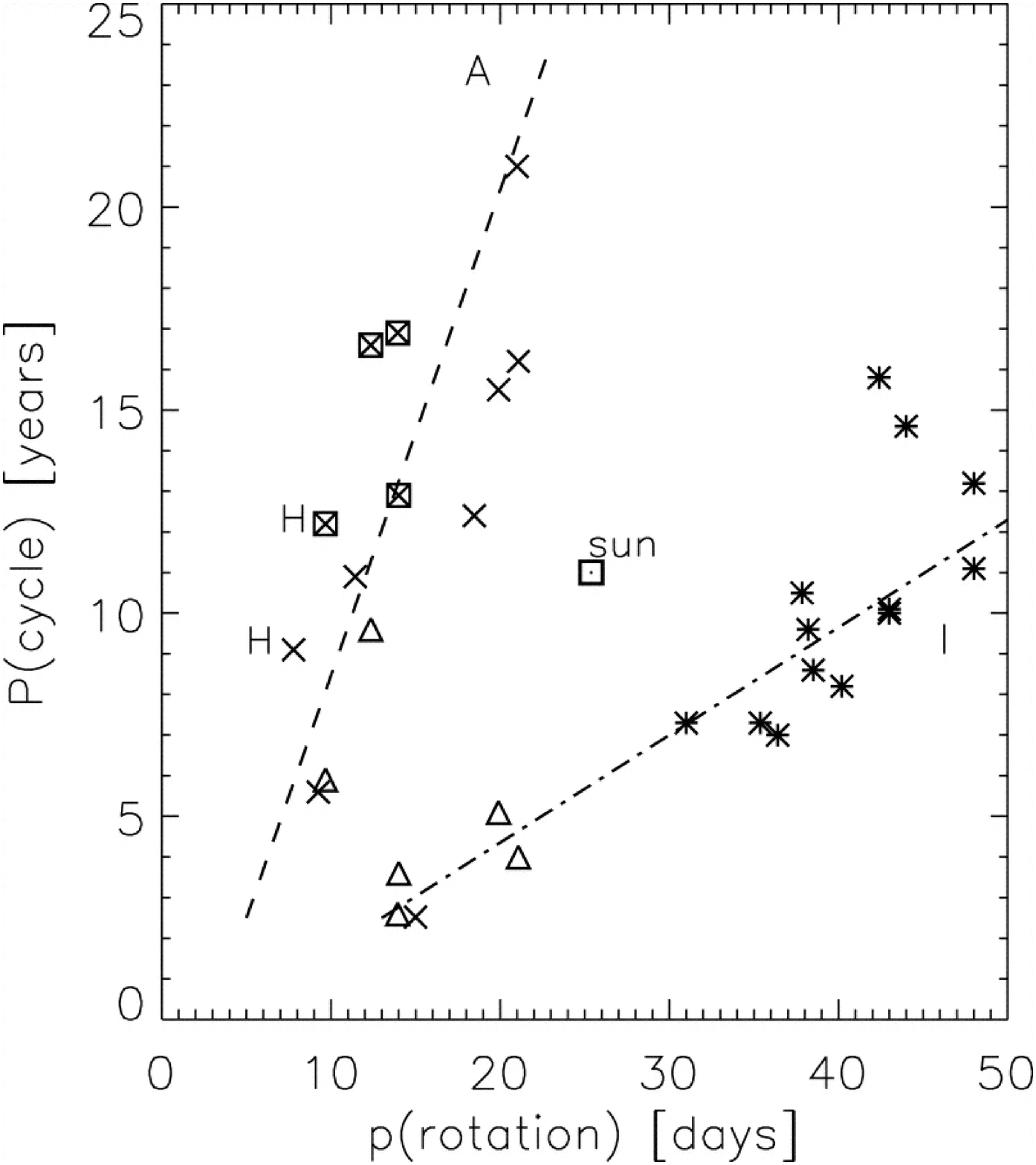} 
 \caption{Periodic behavior derived from the Ca II S index data
(B\"ohm-Vitense 2007).
H indicates Hyades group stars, A and I active and inactive
sequences.
Squares show stars with B -V $<$ 0.62. 
Triangles indicate secondary periods for stars on the A sequence.
}
  \label{fig:bohm}
\end{center}
\end{figure}

Another question often asked is, ``where is a star entering/ exiting
a Maunder Minimum-like phase?''  If the sunspot record of the last 400
years is typical we should see perhaps 1 in 10 cycling stars evolving
into/ out of such grand minima.  This is where the 45 year span of
synoptic measurements is of prime importance- the probability of not
finding such a star dramatically decreases with increasing observing
time spans, as one can become more and more confident that a star has
resumed or has stopped cycling, as the Sun appeared to to in 1745 and
1645 respectively.  The ensemble of solar-like stars cannot address
this point, since we do not really know what fraction of ``flat
activity'' stars truly correspond to Sun-like stars, and not
sub-giants for example Wright (2004).  One clue
might be the observed
anti-correlation of photometric ($b+y$ Stromgren colors) 
and Ca II S indices for inactive stars as well as the Sun.  
In the Sun, 
the brightening due to 
plage emission seems to dominate over the sunspot darkening, at least
when viewed from near the solar equator.  
Hall (2008) discusses one candidate star, HD 140538, in which
the correlation appears to have changed sign from negative to positive
as the star has become
more
active, appearing to begin cycling.  
However, more data are really needed to see if this is indeed
a  star emerging from a ``grand minimum''.

The  sub-convection zone
shear layer (``tachocline'') in the Sun is commonly
assumed to be an essential component of the solar dynamo, although this has
been questioned solely on the basis of solar observations (Spruit 2011).  
The question then arises,  
can cycles exist in fully convective stars?  Well, if the proposal of 
B\"ohm-Vitense (2007) survives further scrutiny, then a tachocline
is not a pre-requisite for a dynamo that can produce
solar-like sunspot cycles.   
In 2007, Cincunegui and colleagues 
tentatively identified a ``cycle'' with period 442 days in
the fully convective star Proxima Cen dM5.5e.  
\nocite{Cincunegui+Diaz+Mauas2007}

\section{Asteroseismology through Kepler}

\label{sec:seismology}
Asteroseismic signals complement the chromospheric and coronal signatures 
discussed above.
The high degree of correlation between the solar magnetic cycle as
seen in the varying 10.7 cm radio emission, and the shifts in acoustic
oscillation frequencies, shown the right hand panel of
Fig.~\protect\ref{fig:morgen}, indicates that asteroseismology has the
capability to reveal solar-like signatures of the solar sunspot cycle
in global oscillations of stars. Both were derived from solar
integrated light data such as might be obtained from more distant stars. Indeed,
asteroseismic observing missions, notably the {\it Kepler} mission, are 
beginning to make this capability a reality. 

Launched in March 2009, Kepler is a NASA mission which is staring at a star 
field in the constellation of Cygnus to look for planets orbiting around 
other stars.  But Kepler is also permitting high-precision asteroseismology 
of the stars within its view.
The broad sweep of asteroseismology achieved in the early phase of the mission 
has been summarized by Chistensen-Dalsgaard \& Thompson (2011).

Stellar activity affects not only the frequencies of the oscillation modes. In
the Sun, the activity cycle and the p-mode {\it amplitudes} are observed
to be anti-correlated. The same anti-correlation is found in the CoRoT star
HD49933. Kepler promises to enable fundamental contributions to the 
understanding of stellar activity by exploiting two complementary aspects of 
the precision photometry. The first aspect is the modulation by star spots, 
which reveals surface activity and surface rotation rate (including differential
rotation). The second aspect is asteroseismology, which reveals 
internal rotation, differential rotation, and internal structure. Observed 
over an extended period, both can reveal stellar cycles. These ideas have been
explored in a 
preliminary analysis of Kepler target stars by Garcia et al.  (2011). 
The initial 
results are very promising, with more than 100 stars already observed 
with rotation periods below the 
10-day period being revealed by these methods. Moreover, the asteroseismic 
analysis has enabled the masses and radii of these stars to be determined. 
However, there are also challenges for this approach, since on the one hand
the stars have to be active enough for starspots to be present and to create
a robust modulation signal of the integrated stellar light, and yet on the 
other hand the modes have still to be of sufficiently large magnitude that the 
asteroseismology remains feasible. A further complication revealed already by
CoRoT is that in many solar-like oscillators the mode lifetimes are rather 
shorter than in the Sun: shorter lifetimes means broader p-mode peaks in the
power spectrum, which makes the measurement of internal rotation more
difficult. But the 
continued study of stars with fast surface rotation is very promising. 

\section{Comparative minima: prospects}

In our short narrative we have suggested that we are only beginning
to examine the magnetic minima of stars for comparison with 
the recent extended solar minimum, and episodes
like the Maunder Minimum.  We can say that the Sun lies at an overall low level of
magnetic activity for a star of its spectral type, but its activity
appears normal for a star of its age.    We have insufficiently long
stellar
time series to understand if these unusual episodes of minimal solar
activity have counterparts in stars (do stars have Maunder Minima?), 
and so we cannot really tell yet
if the Sun's documented variations are in any way unusual. There are
speculations, based upon stars similar to the Sun, and based on the 
recent extended solar minimum, that the Maunder Minimum was a time of
significant  small-scale magnetic activity (Judge \& Saar 2007,
 Schrijver et al.  2011).  It does seem clear, however, from cosmogenic isotope
records, that the Maunder Minimum was a period when the global solar 
field
reversal continued, even in the absence of a strong sunspot count 
(Beer, Tobias \& Weiss 1998).   

How then are we to make progress in this area?  At least three lines
of attack seem worthwhile: (1) We must continue to get much more
``boring'' data, monitoring the photometric and Ca~II emission for decades
into the future: we will test if stars can be found, for example, 
entering/exiting grand minima. (2) Asteroseismology with Kepler and
other experiments will clarify the evolutionary states of large numbers of
stars. The evolution of magnetism in stellar samples will be 
set more quantitatively than is at present possible.  (3) We must 
observe in detail those stars of special interest to our
Sun.  
The star 18 Sco is
the closest ``solar twin'', and the two G2 V stars of the 16 Cygni 
system have recently been studied asteroseismically
(Metcalfe et al.  2012), being two of the brightest targets in the
Kepler field of view.  The two stars are 6.8 Gyr old stars just
slightly more massive than the Sun itself.  The combination of
asteroseismic determinations of mass, age etc., with continued careful
monitoring of Ca~II and other indices presents us with a powerful tool
for probing the magnetic--field evolution of the Sun and stars.

\begin{discussion}

\discuss{Linsky}{There is a new diagnostic, the far ultraviolet continuum emission observed by Hubble COS. We find that active solar-mass dwarf stars have FUV continuum fluxes very similar to bright solar faculae and inactive solar-mass stars have FUV fluxes similar to centers of solar granules. There is a paper by Linsky et al. (2011) now available in Astroph.}  

\discuss{Thompson}{Thank you for pointing this out.}

\discuss{Giampapa}{You mentioned that several hundred solar-type stars have detected p-mode oscillations. Why haven't all solar-type stars in the Kepler sample shown p-mode oscillations?}

\discuss{Thompson}{First I would point out that "solar-type" doesn't mean "solar-twin", for a solar-twin star I would certainly expect an oscillation spectrum essentially like the Sun. Secondly, the solar-type stars without detected oscillations may be oscillating but at an amplitude below that which even Kepler can detect.}

\discuss{Elsworth}{Also, as you pointed out in your talk, activity suppresses p-mode amplitudes. So, active stars in the sample may have unobserved p-modes.}  

\end{discussion}

\end{document}